\documentclass[aps,prl,preprint,superscriptaddress]{revtex4-1}

\usepackage{graphicx}
\usepackage{hyperref}
\usepackage{color}
\usepackage[normalem]{ulem}
\usepackage{lineno}

\begin{document}

\title{Kondo scenario of the $\gamma$-$\alpha$ phase transition in single crystalline cerium thin films}

\author{Xie-Gang Zhu$^\star$}
\affiliation{Science and Technology on Surface Physics and Chemistry Laboratory, Jiangyou 621908, Sichuan, China}

\author{Yu Liu$^\star$}
\affiliation{Laboratory of Computational Physics, Institute of Applied Physics and Computational Mathematics, Beijing 100088, China}
\affiliation{Software Center for High Performance Numerical Simulation,
China Academy of Engineering Physics, Beijing 100088, China}

\author{Ya-Wen Zhao$^\star$}
\affiliation{Institute of Materials, China Academy of Engineering Physics, Jiangyou 621908, Sichuan, China}

\author{Yue-Chao Wang}
\affiliation{Laboratory of Computational Physics, Institute of Applied Physics and Computational Mathematics, Beijing 100088, China}

\author{Yun Zhang}
\affiliation{Science and Technology on Surface Physics and Chemistry Laboratory, Jiangyou 621908, Sichuan, China}

\author{Chao Lu}
\affiliation{Institute of Materials, China Academy of Engineering Physics, Jiangyou 621908, Sichuan, China}

\author{Yu Duan}
\affiliation{Science and Technology on Surface Physics and Chemistry Laboratory, Jiangyou 621908, Sichuan, China}

\author{Dong-Hua Xie}
\affiliation{Institute of Materials, China Academy of Engineering Physics, Jiangyou 621908, Sichuan, China}

\author{Wei Feng}
\affiliation{Science and Technology on Surface Physics and Chemistry Laboratory, Jiangyou 621908, Sichuan, China}

\author{Dan Jian}
\affiliation{Science and Technology on Surface Physics and Chemistry Laboratory, Jiangyou 621908, Sichuan, China}
\affiliation{Laboratory of Computational Physics, Institute of Applied Physics and Computational Mathematics, Beijing 100088, China}

\author{Yong-Huan Wang}
\affiliation{Science and Technology on Surface Physics and Chemistry Laboratory, Jiangyou 621908, Sichuan, China}

\author{Shi-Yong Tan}
\affiliation{Science and Technology on Surface Physics and Chemistry Laboratory, Jiangyou 621908, Sichuan, China}

\author{Qin Liu}
\affiliation{Science and Technology on Surface Physics and Chemistry Laboratory, Jiangyou 621908, Sichuan, China}

\author{Wen Zhang}
\affiliation{Science and Technology on Surface Physics and Chemistry Laboratory, Jiangyou 621908, Sichuan, China}
\affiliation{Sate Key Laboratory of Environmental-Friendly Energy Materials, Southwest University of Science and Technology, Mianyang 621010, Sichuan, China}

\author{Yi Liu}
\affiliation{Science and Technology on Surface Physics and Chemistry Laboratory, Jiangyou 621908, Sichuan, China}

\author{Li-Zhu Luo}
\affiliation{Science and Technology on Surface Physics and Chemistry Laboratory, Jiangyou 621908, Sichuan, China}

\author{Xue-Bing Luo}
\affiliation{Science and Technology on Surface Physics and Chemistry Laboratory, Jiangyou 621908, Sichuan, China}

\author{Qiu-Yun Chen}
\email{sheqiuyun@126.com}
\affiliation{Science and Technology on Surface Physics and Chemistry Laboratory, Jiangyou 621908, Sichuan, China}

\author{Hai-Feng Song}
\email{song\_haifeng@iapcm.ac.cn}
\affiliation{Laboratory of Computational Physics, Institute of Applied Physics and Computational Mathematics, Beijing 100088, China}
\affiliation{Software Center for High Performance Numerical Simulation,
China Academy of Engineering Physics,
Beijing 100088, China}

\author{Xin-Chun Lai}
\email{laixinchun@caep.cn}
\affiliation{Science and Technology on Surface Physics and Chemistry Laboratory, Jiangyou 621908, Sichuan, China}

\date{\today}

\begin{abstract}
The physical mechanism driving the $\gamma$-$\alpha$ phase transition of face-centre-cubic (fcc) cerium (Ce) remains controversial until now. In this work, high quality single crystalline fcc-Ce thin films were grown on Graphene/6$H$-SiC(0001) substrate, and explored by XRD and ARPES measurement. XRD spectra showed a clear $\gamma$-$\alpha$ phase transition at $T_{\gamma-\alpha}\approx$ 50 K, which is retarded by strain effect from substrate comparing with $T_{\gamma-\alpha}$ (about 140 K) of the bulk Ce metal. However, APRES spectra did not show any signature of $\alpha$-phase emerging in the surface-layer from 300 K to 17 K, which implied that $\alpha$-phase might form at the bulk-layer of our Ce thin films. Besides, an evident Kondo dip near Fermi energy was observed in the APRES spectrum at 80 K, indicting the formation of Kondo singlet states in $\gamma$-Ce. Furthermore, the DFT+DMFT calculations were performed to simulate the electronic structures and the theoretical spectral functions agreed well with the experimental ARPES spectra. In $\gamma$-Ce, the behavior of the self-energy's imaginary part at low frequency not only confirmed that the Kondo singlet states emerged at $T_{\rm KS} \geq 80$ K, but also implied that they became coherent states at a lower characteristic temperature ($T_{\rm coh}\sim 40$ K) due to the indirect RKKY interaction among $f$-$f$ electrons. Besides, $T_{\rm coh}$ from the theoretical simulation was close to $T_{\gamma-\alpha}$ from the XRD spectra. These issues suggested that the Kondo scenario might play an important role in the $\gamma$-$\alpha$ phase transition of cerium thin films.
\end{abstract}

\maketitle

\section{Introduction}

Cerium is among one of the most amazing elements across the periodic table, due to its complex phase diagram in which up to seven allotropic phases ($\gamma$, $\beta$, $\alpha$, $\alpha^\prime$, $\alpha^{\prime\prime}$, $\eta$ and $\delta$) could be realized in a modest pressure and temperature range \cite{Alexander:2012aa,Tsiok:2001aa,Schiwek:2002aa,Lipp:2008aa,Tian:2015aa,James:1952aa,Burgardt:1976aa}. And the most intriguing and mysterious part of the phase diagram of Ce is the $\gamma$-$\alpha$ phase transition which involves a huge volume change (up to 16.5\%) \cite{Wilkinson:1961aa}. Great efforts have been devoted to understand the underlying physical mechanism of this unusual phenomenon, both from theoretical and experimental points of view. As for the theoretical explanation, historically, there were several versions: (1) the promotional model, in which it was believed that the $f$-electron of Ce turned from localized (core-level like) to itinerant (valence-like) upon the $\gamma$ to $\alpha$ transition \cite{Coqblin:1968aa,Ramirez:1971aa,Hirst:1974aa}. Therefor the electronic configuration changed from $4f^1[5d6s^2]^{3+}$ to $[4f5d6s^2]^{4+}$. However, this argument was proved to be inconsistent with positron-annihilation experiments \cite{Gustafson:1969aa,Gempel:1972aa}, Compton scattering \cite{Kornstadt:1980aa} and inelastic neutron scattering \cite{Murani:2002aa,Murani:2005aa}, as almost no obvious change in the 4$f$ occupancy was observed. (2) the Mott transition model \cite{Johansson:1974ab}, in which there was no significant modification of the 4$f$ occupancy across the transition. The ratio of on-site Coulomb repulsion energy $U$ and 4$f$ band width $W$, \emph{i.e.}, $U/W$, was the key parameter for this model. It was believed that $U$ and $W$ could be deduced from photoemission (PES) and inverse photoemission (IPES) experiments, and the value of $U/W$ underwent obvious change across the $\alpha$ to $\gamma$ phase transition \cite{Johansson:1974ab}. However, the determination of $U$ and $W$ was unambiguous and challenging. (3) the Kondo Volume Collapse (KVC) model\cite{Allen:1982aa,Allen:1992aa}, in which the Kondo hybridization ($J_{\rm eff}$) between the 4$f$-electron and the conducting ($c$) valence electrons was assumed to vary with volume between the $\gamma$ and $\alpha$ phases. The original conclusion was that the rapid (exponential) change in the Kondo energy ($k_BT_{\rm KS}$) as a function of $J_{\rm eff}$ drove the volume collapse phenomenon. It should be stressed that the key ingredient in the KVC model is the $f$-$c$ hybridization, which means $f$-electrons become indirectly bonding due to the Ruderman-Kittel-Kasuya-Yosida (RKKY) interaction\cite{Zhang:2000aa,Burdin:2001aa,Reinert:2001aa,Yang:2008aa}. Therefore, a direct and clear observation of the $f$-$c$ hybridization effect is inevitable for the understanding of the $\gamma$ to $\alpha$ phase transition of Ce metal.

Angle resolved photoemission spectroscopy (ARPES)  has been proved to be a powerful tool for the direct characterization of electronic structures of materials. High quality single crystal with well-oriented crystalline surface is a necessity for ARPES experiments. As for Ce, it is quite challenging to meet the above criteria, due to its highly chemical reactive nature. Even though macroscopic scale (up to several millimeters) sized bulk Ce single crystal had been successfully synthesized and characterized by neutron scattering techniques \cite{Stassis:1979aa,Stassis:1982aa,Krisch:2011aa}, no high quality ARPES spectra on those bulk samples were ever reported up to now.  Dispersive band structures were only realized on Ce thin films grown on W(110) substrate \cite{Weschke:1998aa,Schiller:2003aa,Chen:2018aa}. However, those works on photoemission demonstrated discrepancies on the $4f$ electronic structures due to their different interpretations of the crystal structure of Ce films grown on W(110). Information on the evolution of the bulk crystalline structure, especially the atomic stacking sequences perpendicular to the thin film surface, was not properly stated \cite{Weschke:1998aa,Schiller:2003aa,Morimoto:2009aa,Chen:2018aa}.

To make a one to one correlation of 4$f$-electron properties and the crystalline structures of Ce metal, here in our work, we have synthesized high quality single crystal Ce thin films on graphene substrate grown on 6$H$-SiC(0001). We chose this substrate based on the lattice match condition and inertness of graphene (detailed discusstion could be found in the Supplementary materials). By a combination of high-resolution STEM at room temperature and varied temperature X-ray diffraction (XRD) down to 15 K, we found that the \emph{as-grown} Ce thin films surprisingly sustained the $\gamma$-phase from 300 K down to around 50 K, which is controversial to commonly accepted phase transition temperature (141$\pm$10 K) of Ce metal at ambient pressure \cite{Koskenmaki:1978ab}. And the $\alpha$-phase emerged in the thin film around 50 K, but the $\gamma$-$\alpha$ transformation did not completely finish even at 15 K, resulting in a mixture of $\alpha$- and $\gamma$-Ce. The \emph{in-situ} ARPES experiments were taken at various temperatures, and a strong hybridization was observed for the $\gamma$-Ce at low temperature. Besides, theoretical calculations by combining the density functional theory and the dynamical mean-field theory (DFT+DMFT) methods were carried out to simulate electronic structures of fcc-Ce upon cooling\cite{Haule2010,Blaha:2001aa}. The calculated spectral functions agreed well with the experimental ARPES spectra. Moreover, the characteristic temperatures of Kondo singlet states formation $T_{\rm KS}$ and the formation of coherent heavy quasi-particle states $T_{\rm coh}$ were simulated numerically, which correspond to the local $f$-$c$ Kondo hybridization and the addtional non-local $f$-$f$ RKKY interaction, respectively \cite{Zhang:2000aa,Burdin:2001aa,Reinert:2001aa,Yang:2008aa}.

\section{Results}

\begin{figure*}
\centering
\includegraphics[width=1.0\textwidth]{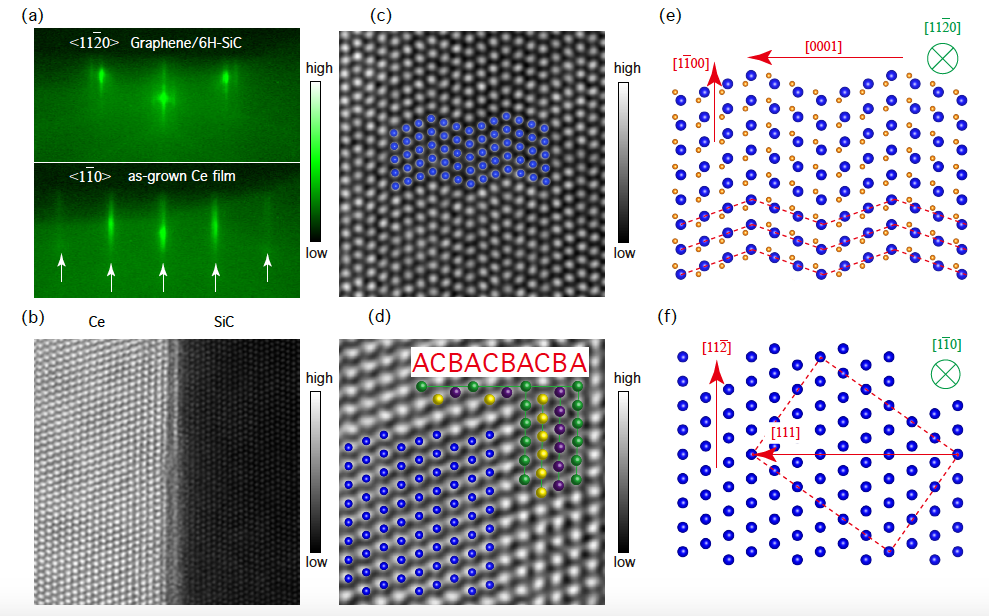}
\caption{(Color online) \textbf{RHEED and HRSTEM results.} (a) RHEED patterns of Graphene/6$H$-SiC substrate and \emph{as-grown} Ce thin film, with incident electron beam along $\left<11\bar{2}0\right>$ direction of the SBZ of 6$H$-SiC(0001); (b) a typical HRSTEM image of the Ce thin film together with the SiC substrate; (c) and (d) arrangement of the Si atoms and Ce atoms; (e) and (f) simulations of the atomic stacking sequences of SiC substrate and the Ce thin films, respectively.}
\label{CrysStruc}
\end{figure*}

\textbf{Crystal structure characterization}.
The reflected high-energy electron diffraction (RHEED) patterns of the Graphene/6$H$-SiC substrate and the \emph{as-grown} Ce thin film are shown in Fig. \ref{CrysStruc}(a). The incident electron beam was along the $\left<11\bar{2}0\right>$ direction, \emph{i.e.}, $\bar{\Gamma}\bar{\text{K}}$ in the surface Brillouin zone (SBZ) of the 6$H$-SiC(0001) substrate. Five streaks are indicated by white arrows on the RHEED pattern of Ce thin film. And this five-streak feature persisted all through the film growth and there was no extra streak, which implies the single crystalline nature of the thin film. It should be noted that the stacking sequence along the $c$-axis of $\beta$-Ce is -ABACABAC-, while it is -ABCABC- perpendicular to the $\gamma$-Ce(111) surface. As the stacking sequence of the first two atomic layers are identical for the two phases, the question arises in which phase the Ce atoms condensed on the substrate. To clarify this, high-resolution transmission electron microscopy (HRSTEM) was used to probe the stacking sequence perpendicular to the sample surface.

A Ce thin film with thickness of 200 nm was used to prepare the sample for HRSTEM experiments. A part of the thin film together with the substrate was cut by focused ion beam (FIB). Fig. \ref{CrysStruc}(b) is a typical HRSTEM image acquired such that the incident electron beam is parallel to the $\left[11\bar{2}0\right]_{6H\text{-SiC}}$/$\left[1\bar{1}0\right]_{\text{Ce}}$ direction.
The atomic arrays of the SiC substrate and the Ce film are extracted and displayed in Fig. \ref{CrysStruc}(c) and (d), respectively. Due to the low atomic mass of Carbon atoms, only the Silicon atoms are revealed in the HRSTEM pattern. The arrangement of the Si atom arrays in Fig. \ref{CrysStruc}(c) could be well reproduced by simple simulation from the well-known crystal structure of 6$H$-SiC\cite{Bauer:1998aa}, as shown in Fig. \ref{CrysStruc}(e). In the simulation in Fig. \ref{CrysStruc}(e), the projection is along $\left[11\bar{2}0\right]$ direction and the left and upward arrows denote the $\left[0001\right]$ and $\left[1\bar{1}00\right]$ directions, respectively. By omitting the Carbon atoms (golden spheres), the zig-zag chains well reproduce the patterns in Fig. \ref{CrysStruc}(c). Furthermore, the atom column in Fig. \ref{CrysStruc}(d) could be grouped in three types, as marked by green (A), purple (B) and blue (C) spheres. They arrange in perfect -ABCABC- sequence, and are reproduced by the simulation of the $\gamma$-phase Ce with projection along $\left[1\bar{1}0\right]$ direction and the column array along $\left[11\bar{2}\right]$ direction. It should be stressed that there was no oxidation of the Ce by checking the EELS signal. From the HRSTEM images, the orientation relationship (OR) between Ce and 6$H$-SiC substrate is expressed as
$\left[111\right]_{\text{Ce}}$//$\left[0001\right]_{6H\text{-SiC}}$. Therefore, our HRSTEM results indicate that the \emph{as-grown} Ce thin film is in the $\gamma$-phase, \emph{i.e.}, having the FCC structure.

\begin{figure*}
\centering
\includegraphics[width=1.0\textwidth]{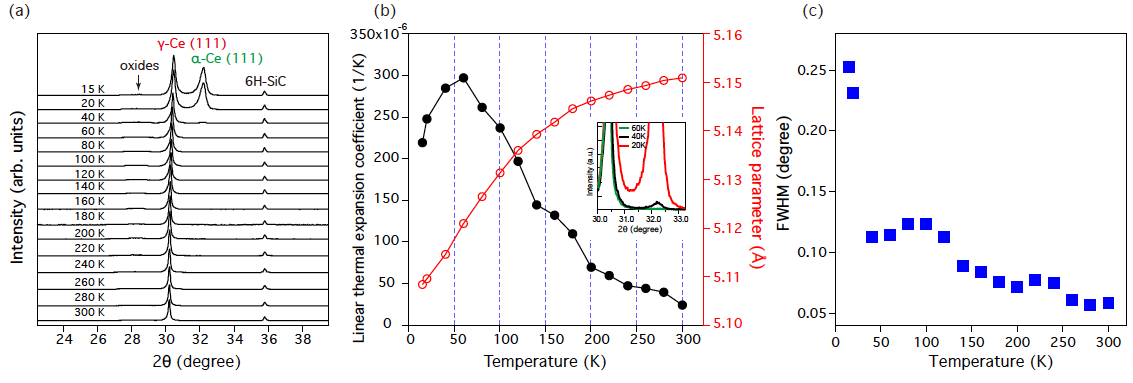}
\caption{(Color online) \textbf{Varied temperature X-Ray Diffraction of a 200 nm thick Ce thin film.} (a) Evolution of the (111) diffraction peaks for $\gamma$ and $\alpha$-Ce, together with the oxide hump (indicated by black arrow) and the 6$H$-SiC substrate, as the sample temperature decreased from 300 K to 15 K. (b) The lattice constant of $\gamma$-Ce and its linear thermal expansion coefficient $\alpha_L$ (defined in the text) versus the temperature were extracted from (a). The inset enlarges the (111) diffraction peaks of the Ce thin film at 60 K, 40 K and 20 K, demonstrating the emergence of $\alpha$-phase Ce. (c) FWHMs of the $\gamma$-Ce(111) peaks at various temperatures.}
\label{vt.xrd}
\end{figure*}

To clarify the lattice structure evolution of the Ce film versus the temperature, \emph{ex-situ} XRD was used to characterize the lattice structure of the Ce film. A 200 nm Ce thin film sample was cooled from 300 K to 15 K and the XRD patterns were collected every 20 K except for the lowest temperature (15 K) as shown in Fig. \ref{vt.xrd}(a). There are several features among the 2$\theta$ range of 24$^\circ$ to 38$^\circ$: a broad hump near 28$^\circ$ which belongs to the oxidized Ce at the sample surface, as oxidation was inevitable during the sample transfer for \emph{ex-situ} XRD experiments; persistent (111) peak of $\gamma$-Ce around 30$^\circ$ at all temperatures; emergence of the (111) peak of $\alpha$-Ce at low temperature; unchanged (0006) peak of the 6$H$-SiC substrate. It is quite interesting that there was no sign of $\beta$-Ce during the phase transition. Fig. \ref{vt.xrd}(b) shows the evolution of the lattice constant ($L$) of $\gamma$-Ce in Fig. \ref{vt.xrd}(a) and its linear thermal expansion coefficient $\alpha_L$ (defined as $\frac{\Delta L}{L\cdot\Delta T}$) versus the temperature. The inset in Fig. \ref{vt.xrd}(b) demonstrates the (111) diffraction peaks at 20 K, 40 K and 60 K, respectively. The $\alpha$-Ce (111) diffraction peak emerged below 60 K, which offers a fingerprint for the $\gamma$-$\alpha$ phase transition. And the phase transition induced a crossover of the $\alpha_L$ curve around 60 K.

From our XRD results, the $\gamma$-$\alpha$ phase transition temperature in thin film lagged far behind the bulk crystal (141$\pm$10 K), which can be attributed to the strain effect in our samples with the lattice match of the in-plane lattice parameters of the substrate and the (111) surface of $\gamma$-Ce. We qualitatively studied the strain effect through analyzing the full-width at half maximum (FWHM) of the XRD diffraction peaks at various temperatures, as peak width contains rich information on the crystallite size, defect (strain, disorder, defects) of the studied materials. The FWHMs of the $\gamma$-Ce(111) peaks at various temperature were extracted by fitting with Lorentzian line-shape which was convolved with a Gaussian-type instrumental broadening (0.05$^\circ$), as shown in Fig. \ref{vt.xrd}(c).The peak width showed no obviously change from 300 K to 180 K (0.068 $\pm$ 0.009$^\circ$), and started to increase to a rather flat plateau between 140 K and 40 K (0.117 $\pm$ 0.004$^\circ$) as temperature further decreased. Abrupt change in the peak width happened below 40 K, indicating the phase transition posed obviously effects on the crystalline status of the thin films, which is consistent with the emergence of $\alpha$-Ce(111) XRD peak at 40 K(inset of Fig. \ref{vt.xrd}(b)).

The phenomena observed above indicate that the thin film tended to transform from $\gamma$-phase to $\alpha$-phase below 140 K, which is consistent with the phase transition temperature of bulk crystals in the literatures. However, the phase transition did not happen until around 50 K, as a result of the existence of the substrate/thin film interface. It should be noted that, even with the existence of strong interface effect, the lattice of the thin film continued to shrink as temperature decreased, indicating the competition of temperature and interface effects on the phase transition of the thin film. The abrupt change in the peak width at 40 K indicates the interface effects were overwhelmed.

Additionally, we have done Low Energy Electron Diffraction (LEED) experiments on Ce thin films at 80 K (see the Supplementary materials for detail). We found that the in-plane lattice parameter of Ce thin film at 80 K is 3.60 $\pm$ 0.02 \AA, and this value is comparable to the VT-XRD results at 80 K, \emph{i.e.}, $\approx$ 3.62 \AA. Therefore, the surface of Ce thin film remains in the $\gamma$-phase at 80 K. The persistent existence of pure $\gamma$-Ce down to 60 K makes it possible to study the evolution of the 4$f$ electronic properties versus temperature.

\begin{figure*}
\centering
\includegraphics[width=1.0\textwidth]{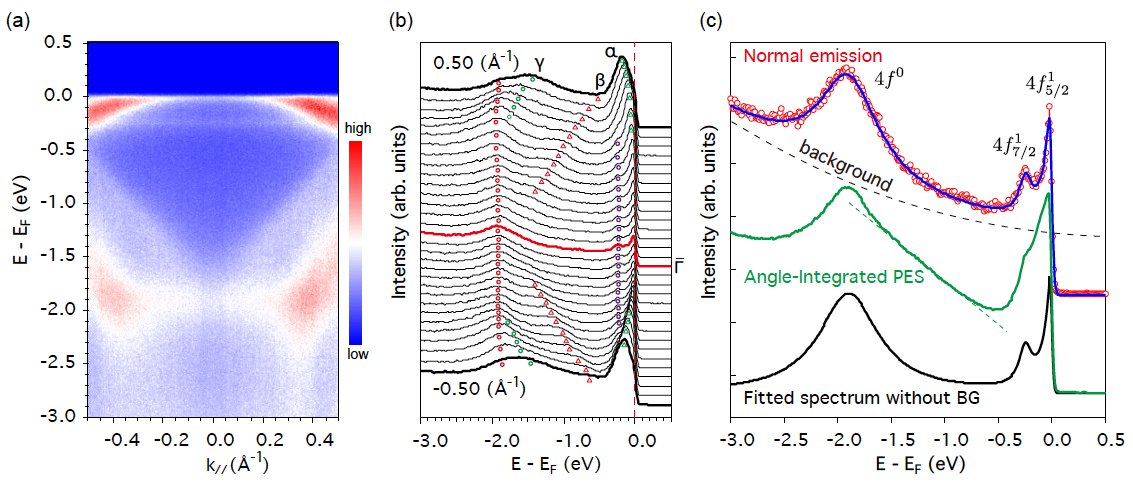}
\caption{(Color online) \textbf{Band structures and EDCs fitting results.} (a) Large-scale ARPES spectrum of Ce thin film at T = 80 K. (b) Energy distribution curves (EDCs) of (a), demonstrating the three conduction bands $\alpha$, $\beta$ and $\gamma$, together with $4f^0$, $4f^1_{5/2}$ and $4f^1_{7/2}$ levels. (c) Normal emission and angle-integrated photoemission spectra of (a).}
\label{lsPES}
\end{figure*}

\textbf{Photoemission studies}. An \emph{as-grown} Ce thin film sample with thickness of 20 nm was transfered \emph{in-situ} into the ARPES chamber under UHV within 5 minutes after growth. Because of the high chemical reactivity of Ce, the duration of the PES measurement was restricted within one hour after the sample temperature reached the desired value. ARPES spectrum at 80 K is shown in Fig. \ref{lsPES}(a), with $k_{//}$ along $\bar{\Gamma}\bar{\rm K}$ in the surface Brillouin zone. Strongly dispersive electronic bands indicate the high quality of our single crystal film. There are three predominant non-dispersive features in the spectra. The broad flat band at a binding energy of $\approx$ 2.0 eV below $E_{\rm F}$ corresponds to the $4f^1\rightarrow4f^0$ ionization peak, and is labeled as 4$f^0$. The other two flat bands situate near Fermi level and at $\approx$ 250 meV below Fermi level, corresponding to the so-called Kondo resonance and its spin-orbit coupling replica \cite{Patthey:1985aa}, and are traditionally labeled as $4f^1_{5/2}$ and $4f^1_{7/2}$, respectively. Those three features could be identified more specifically from the energy distribution curves (EDCs) as shown in Fig. \ref{lsPES}(b). The 4$f^0$ and 4$f^1_{7/2}$ bands are marked by red and purple circles, respectively. And the 4$f^1_{5/2}$ signal dominates the intensity near Fermi level. Apart from those 4$f$ features, there exist two conduction bands between 4$f^0$ and 4$f^1_{7/2}$, labeled as $\beta$- and $\gamma$-band, respectively. Here we also present the normal emission (NE) and angle-integrated photoemission (AIPES) spectra of Fig. \ref{lsPES}(a) in Fig. \ref{lsPES}(c). The normal emission and AIPES spectra were extracted from Fig. \ref{lsPES}(a) by integrating the EDCs within $\pm 0.02$ $\text{\AA}^{-1}$ and $\pm 0.50$ $\text{\AA}^{-1}$ around the $\bar{\Gamma}$ point, respectively. At the Fermi energy ($E_{\rm F}$), AIPES shows a quite broad peak, which extends to about -0.5 eV, corresponding to the the band width of the $\alpha$ band. The 4$f^1_{7/2}$ peak is considerably suppressed in the AIPES and could be barely figured out as a little hump near -200 meV. What is more, the valence bands between 4$f^1_{7/2}$ and 4$f^0$, \emph{i.e.}, $\beta$ and $\gamma$ bands, contribute a nearly linear density of states (DOS) to AIPES, as indicated by the dashed green line on the AIPES spectrum. This linear DOS adds up to the single 4$f^0$ state, resulting in a anisotropic broad 4$f^0$ peak that cannot be fitted by a standard Lorentzian or Gaussian line shape. However, the situation for the NE spectrum is quite different. Sharp 4$f^1_{5/2}$ peak sits near $E_{\rm F}$ and the 4$f^1_{7/2}$ level could be resolved quite well. The NE spectrum could be simulated by a multi-component function,
\begin{equation}
f(\epsilon)=A_0+\left[P_3(\epsilon)+\sum_iL(\epsilon,\epsilon_i, w_i)\right]\bullet[F*G](\epsilon).
\label{fit.func}
\end{equation}
Here, $A_0$ is the overall constant background due to the experimental noise, $P_3(\epsilon)$ is a cubic polynomial that simulates the spectral contributions from valence bands, and Lorentzian lineshape $L(\epsilon,\epsilon_i,w_i)$ centers at $\epsilon_i$ with a full width at half maximum (FWHM) of $w_i$. $[F*G](\epsilon)$ corresponds to the convolution of the Fermi-Dirac distribution and a Gaussian instrumental broadening. The Gaussian broadening was estimated by fitting the AIPES of an amorphous gold sample at 80 K and a broadening of $\approx$ 18.5 meV was obtained. Three Lorentzian peaks are used to fit the normal emission spectrum. The solid blue curve in Fig. \ref{lsPES}(c) is the fitted result which nicely reproduces the NE spectrum. The spectral contribution from 4$f^0$, 4$f^1_{7/2}$ and 4$f^1_{5/2}$ are extracted and demonstrated as black solid lines in Fig. \ref{lsPES}(c). The extracted spectrum resembles the resonance photoemission (PE) results of $\alpha$-Ce in Ref. \onlinecite{Weschke:1991aa} and \onlinecite{Kucherenko:2002aa}, in which the 4$f$ signals of Ce are resonantly enhanced due to the $4d\rightarrow4f$  absorption threshold and the PE cross section of valence electrons are low. Our observed distribution of $4f$ electronic states is quite striking, as the previous VT-XRD experiments showed clearly the Ce thin film retains $\gamma$ phase above 60 K. However, we should keep in mind that, in the Kondo scenario, Kondo effect would manifest itself as the system is cooled near or lower than the Kondo temperature $T_{\rm KS}$, strengthening the hybridization of $f$ states with valence states ($f$-$c$). The intense $4f^1_{5/2}$ peak near $E_{\rm F}$ at 80 K invokes a carefully investigation of the electronic structures close to $E_{\rm F}$ at various temperatures in pursuit of the possible $f$-$c$ hybridization evidence in Ce metal.

\begin{figure*}
\centering
\includegraphics[width=1.0\textwidth]{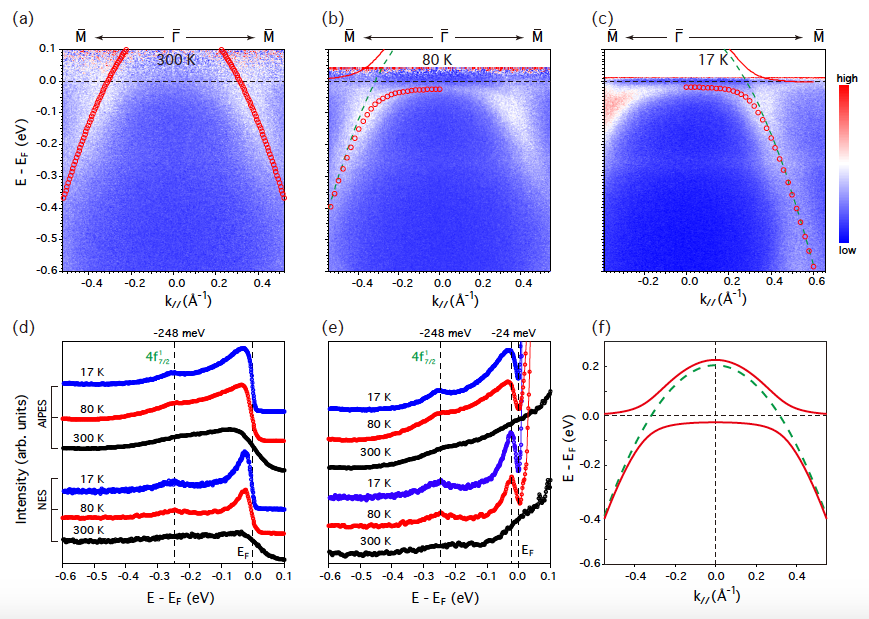}
\caption{(Color online) \textbf{Evolution of the $c$-$f$ hybridization with temperature and the PAM model.} (a)-(c) ARPES spectra of 20 nm thick Ce thin films at 300 K, 80 K, and 17 K, respectively (with RC-FDD correction); (d) and (e) NE and AIPES spectra of (a)-(c), without/with RC-FDD correction; (f) illustration of strong $f$-$c$ hybridization within the PAM framework.}
\label{vtPES}
\end{figure*}

Fig. \ref{vtPES}(a)-(c) show the ARPES spectra collected at 300 K, 80 K and 17 K, within an energy range from 600 meV below $E_{\rm F}$ to 100 meV above $E_{\rm F}$. To have a better view of the fine structures near $E_{\rm F}$, the spectra were divided by the corresponding resolution-convoluted Fermi-Dirac distribution (RC-FDD) \cite{Chen:2017aa}, with an instrumental broadening of 25 meV, 14.5 meV and 10 meV, respectively. Obviously, the spectrum at 300 K has the largest momentum and energy broadening due to the thermal effect at high temperature. The two branches of $\alpha$ band dominate the spectral weight, with almost vanishing 4$f^1$ features, indicating a localized nature of the $4f$ electrons at 300 K. As the temperature dropped to 80 K, the spectral weight of $4f^1_{5/2}$ and $4f^1_{7/2}$ level developed. Meanwhile, the valence bands $\alpha$ dispersed toward $\bar{\Gamma}$ point and merged with $4f^1_{5/2}$ level, demonstrating a strong evidence for the notable $f$-$c$ hybridization, which is the very first one observed in $\gamma$-Ce at low temperature. Further cooling Ce thin film down to 17 K gave no fundamental change in the ARPES spectrum, except that the dispersions are sharper and even the hybridization between $\alpha$ band and $4f^1_{7/2}$ produces noticeable distortion of the $\alpha$ band near -250 meV. We noted that our extracted NES and AIPES, as shown in Fig. \ref{vtPES}(d) and (e), resemble the resonance photoemission results of $\alpha$-Ce in Ref. \cite{Weschke:1991aa} and \cite {Kucherenko:2002aa} in the following way: similarity in the intensity ratio of $4f_1^{5/2}$ and $4f_1^{7/2}$ states. And what is more important, the position of $4f^0$ ionization peak in our work matches very well with that of $\gamma$-phase Ce in Ref. \cite{Patthey:1985aa,Weschke:1991aa,Kucherenko:2002aa} (see detail in the Supplementary materials). Therefore, we were probing $\gamma$-phase Ce on the thin film surfaces in our ARPES experiments. It is worth noting that a ``$\gamma$-$\alpha$'' phase transition of monolayer Ce on W(110) was reported \cite{Gu:1991ab} and a splitting of $4f^0$ ionization peak was revealed in a Ce monolayer on W(110) substrate, which was attributed to the formation of Ce $4f$ band and its hybridization with the valence-band states \cite{Vyalikh:2006aa}. Our current work differs with this work in the sense that we were dealing with rather thick Ce thin films (dozens of nanometers) and focussing on the $f$-$c$ hybridization near $E_{\rm F}$, not at $4f^0$. The photon energy that we used is quite surface sensitive, \emph{i.e.}, it could probe only a few atomic layers on the sample surface. The similarities in the ARPES spectra indicate that the sample surfaces possess the same phase structure at 80 K and 17 K. Taking the VT-XRD experiments in Fig. \ref{vt.xrd} into consideration, we could conclude that the $\gamma$-$\alpha$ phase transition undergoes beneath the sample surface, \emph{i.e.}, starting from the bulk layers of the sample. Nevertheless, this unexpected phenomenon offers us the possibilities for exploring the evolution of the electronic structure of $\gamma$-Ce at low temperature, especially the localized to itinerant transition of $4f$ electrons in Ce.

Usually, in the Kondo scenario for Ce metals and compounds, the Kondo resonance (KR), an enhanced electron density of states (DOS), has a maximum above $E_{\rm F}$. And PES experiments have merely access to the tail the KR below $E_{\rm F}$. Surprisingly, the RC-FDD corrected ARPES spectra at 80 and 17 K revealed a weakly dispersed bands $\approx$ 24 meV below $E_{\rm F}$, sufficiently high in binding energy to be resolved within the instrumental resolution, as could be seen from the NE and AIPES spectra in Fig. \ref{vtPES}(d) and (e). A similar spectral feature at about 21 meV below $E_{\rm F}$ was also observed in the superconducting heavy Fermion compound $\text{CeCu}_2\text{Si}_2$ \cite{Reinert:2001aa}, which arises from the virtual transition from the excited crystal field (CF) splitting to the ground $4f^1$ state and thus has a much lower spectral weight than the corresponding KR peak. The CF splitting in $\gamma$-Ce was estimated to be 5.8 meV by inelastic neutron scattering \cite{Millhouse:1974aa}, far less that 24 meV. Therefore, the spectral feature near 24 meV below $E_{\rm F}$ is a pure demonstration of the ground state hybridized with valence states. To investigate the $k$-dependent $f$-$c$ hybridization, we introduce the phenomenological periodic Anderson model (PAM), which gives the band dispersion describing the hybridization as \cite{Chen:2017aa},
\begin{equation}
E_k^{\pm}=\frac{\epsilon_0+\epsilon(k)\pm\sqrt{(\epsilon_0-\epsilon(k))^2+4|V_k|^2}}{2}.
\label{fit.pam}
\end{equation}
Here $\epsilon_0$ is $4f$ ground state energy, $\epsilon(k)$ is the valence band dispersion at high temperature, and $V_k$ is the renormalized hybridization strength. As shown in Fig. \ref{vtPES}(d) and (e), there is weak non-vanishing spectral weight around -250 meV at 300 K in the NES and AIPES spectra, which corresponds to the $4f_1^{7/2}$ level. We should note that the thermal broadening at 300 K ($\approx$ 26 meV) smeared the $4f_1^{5/2}$ level and gave rise to a continuous upward trend for the NES and AIPES spectra in Fig. \ref{vtPES}(e) from -0.1 to 0.1 eV and above. Here we took the two branches of the dispersive bands at 300 K as unhybridized valence bands, and fitted them by a hole-like parabolic band, and kept its shape fixed for the fittings at low temperature. As the temperature dropped to 80 K, the dispersions below $E_{\rm F}$ could be fitted perfectly by Eq. \ref{fit.pam}, and give $\epsilon_0 = -4.0\pm2.6$ meV, and $V_k=71 \pm 5$ meV, respectively.
The fitting results indicate that the ground state of $\gamma$-Ce, \emph{i.e.}, $4f^1$ has a position just several meV below $E_{\rm F}$. Moreover, as a result of the strong $f$-$c$ hybridization strength $V_k$, the resultant final states have a direct gap (defined as a minimal separation of two bands at the same momemtum) of $\approx$ 140 meV and an indirect gap (defined as the global minimal separation of two bands) of $\approx$ 20 meV, respectively. What is more, the fitting results could reproduce the hybridization behaviors very well at low temperature. This fact indicates that the Fermi crossing of the unhybridized valence bands remained almost the same upon cooling from 300 K to 80 K and 17 K, meaning that we were probing $\gamma$-Ce on the sample surfaces in our ARPES experiments.

On the one hand, from the XRD experiments, pure $\gamma$-phase exists above the phase transition critical temperature $T_{\gamma-\alpha}\approx 50$ K, while $\gamma$- and $\alpha$-phase obviously coexist  below this critical temperature. However, on the other hand, the ARPES spectra above and below $T_{\gamma-\alpha}$ are similar, which means there is just only one phase in the surface layers of the sample in the whole temperature region. Therefore, it can be deduced that below the phase transition critical temperature,  the $\alpha$-phase could emerge in the bulk first, while the $\gamma$-phase remained unchanged in the surface. Although the bare Coulomb interaction strength $U$ and $J$ are similar in the surface and bulk, their effective strength in the surface might be much larger than those in the bulk due to the weaker screening effect from the surface $c$-electrons. This means that the $f$-electrons in the surface are more localized than those in the bulk, which results in a smaller cohesive energy between the Ce atoms in the surface region. Therefore, the environment of the surface favors the survival of $\gamma$-phase. In other words, as the temperature decreases, the $\gamma$-$\alpha$ phase transition happens more easily in the bulk than in the surface. It is similar with the so-called ``surface Kondo breakdown'' used to explain unusual quantum oscillations in the topological Kondo insulator SmB$_6$ \cite{Victor:2015aa, Erten:2016aa, Robert:2016aa}.

\begin{figure*}
\centering
\includegraphics[width=1.0\textwidth]{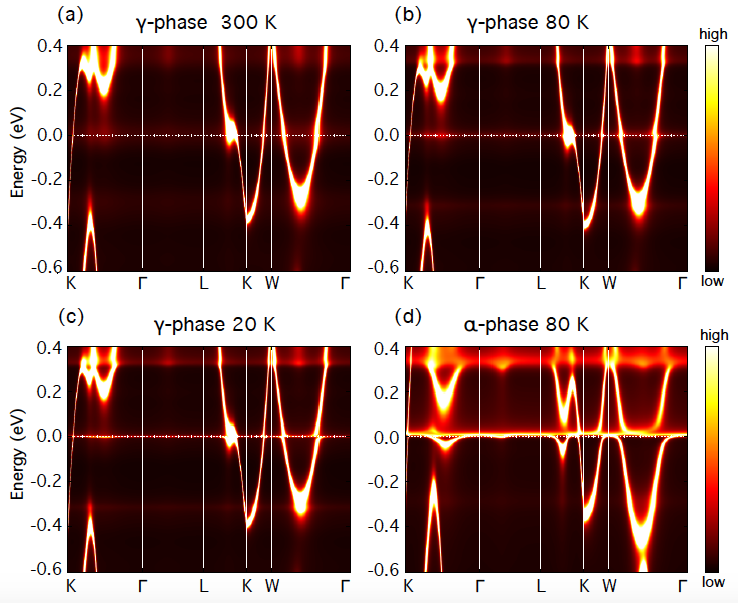}
\caption{(Color online) \textbf{DFT+DMFT calculations.} (a)-(c) Momentum-resolved spectral functions $A(k,\omega)$ along the high-symmetry lines in the Brillouin zone obtained by DFT+DMFT calculations of $\gamma$-Ce at 300 K, 80 K and 20 K. (d) Spectral functions $A(k,\omega)$ of $\alpha$-Ce at 80 K.}
\label{dmft}
\end{figure*}

\textbf{DFT+DMFT calculations}. From our theoretical calculations, the momentum-resolved spectral functions along high symmetry path at 300 K, 80 K and 20 K for $\gamma$-Ce are shown in Fig.~\ref{dmft}(a)-(c). The two $f$ levels, \emph{i.e.}, $4f_1^{5/2}$ and $4f_1^{7/2}$ locate near $E_{\rm F}$ and about 280 meV below $E_{\rm F}$, respectively. And their contribution to the density of states could be found in Supplementary Figure 3, where we could clearly see the non-vanishing $4f^1$ spectral weight even at 300 K. We could observe that the flat hybridization bands emerge at 80 K and become more evident at 20 K, which are not present at 300 K. For comparison, in Fig.~\ref{dmft}(d), we can clearly observe that the effective $f$-$c$ hybridization is much stronger in the $\alpha$-phase at relatively high temperature, \emph{i.e.}, 80 K, such as the flat hybridization bands near $E_{\rm F}$ become more evident and the other low-energy excitation bands shift closer to $E_{\rm F}$. Besides, there exist many other characteristic differences between the two phases, such as two much larger electronic Fermi surfaces appear at $K$-$\Gamma$ and $L$-$K$ path which did not form well at $\gamma$-phase even at 20 K. These agree well with previous theoretical calculation results based on the continuous-time quantum Monte Carlo impurity solver \cite{Chen:2018aa}.

\begin{figure*}
\centering
\includegraphics[width=1.0\textwidth]{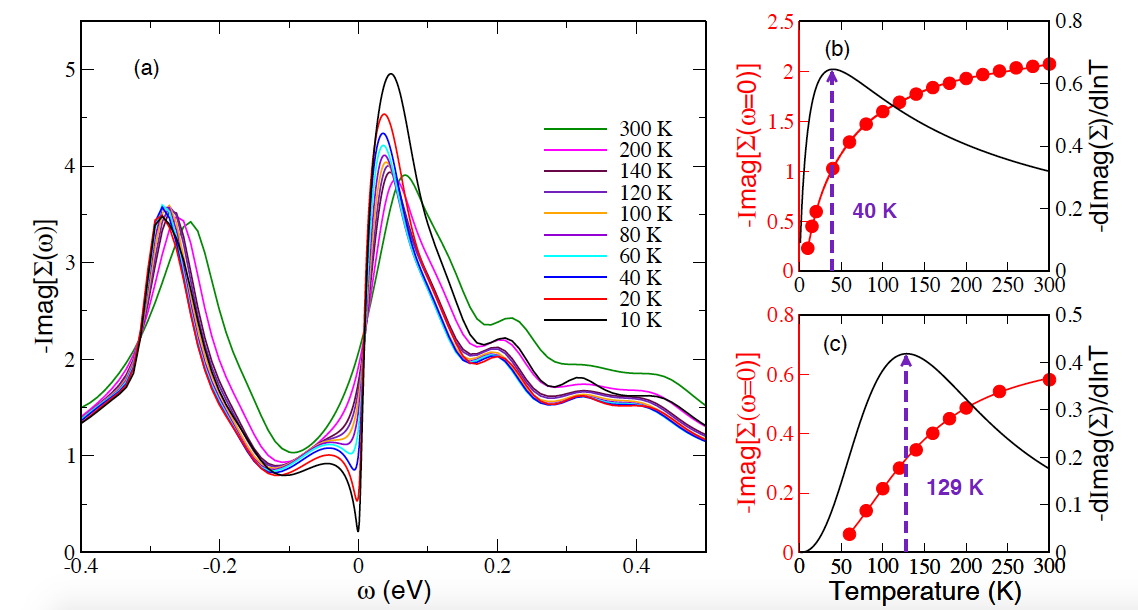}
\caption{(Color online) \textbf{Evolution of the imaginary part of the self-energy versus temperatures.} (a) The magnitude of the imaginary part of the 4$f$ ($J=5/2$) self-energy (-${\rm Im} \Sigma(\omega,T)$) versus energy $\omega$ at different temperatures; (b)-(c) the temperature derivative of -${\rm Im} \Sigma(\omega,T)$ at zero energy ($\omega=0$) for $\gamma$-Ce and $\alpha$-Ce, respectively. The results show a maximum temperature derivative at about 40 K for $\gamma$-Ce and 129 K for $\alpha$-Ce.}
\label{sigma}
\end{figure*}

In order to give a deeper investigation of this system, we need to analysis the self-energy as it encodes all the electronic correlations in DFT+DMFT calculation. The lifetime of quasi-particle could be indirectly reflected by the magnitude of the self-energy's imaginary part,\emph{i.e.}, ${\rm Im} \Sigma(\omega,T)$. Therefore, certain crossover feature could be observed from the behavior of $-{\rm Im} \Sigma(\omega, T)$  varied with the frequency ($\omega$) and temperature ($T$) \cite{Shim:2007aa}.
Since the low-energy excited heavy quasi-particle is mainly composed of the 4$f$-electrons with $J=5/2$, curves of $-{\rm Im} \Sigma_{5/2}(\omega,T)$ versus $\omega$ at different temperatures for $\gamma$-Ce are plotted  in Fig.~\ref{sigma}(a). An evident dip near the static limit ($\omega\rightarrow0$) emerged gradually as cooling below $T_{\rm KS}\approx$ 80 K, representing the formation of the flat hybridization bands due to the localized Kondo hybridization, which can be seen clearly from the derivative of $-{\rm Im} \Sigma_{5/2}(\omega,T)$ with respect to energy $\omega$ in the Supplementary Figure 3(c). It could be also confirmed by our experimental ARPES spectra and theoretical momentum-resolved spectral functions.
The functions $-{\rm Im} \Sigma_{5/2} (\omega=0,T)$ are plotted as red points and fitted by the red lines in Fig.~\ref{sigma}(b)-(c), for $\gamma$ and $\alpha$ phase, respectively. Here we have adopted a four-parameter logistic fitting function, \emph{i.e.}, $y=(A_1-A_2)/[1+(T/T_0)^p]+A_2$ (where $A_1$, $A_2$, $T_0$, and $p$ are the fitting parameters), which is widely used to to elucidate the turning point of the change of rate of a curve versus certain physical quantity. Besides, their logarithmic temperature derivatives $ -{\rm d Im} \Sigma_{5/2} (\omega=0,T)/ {\rm d ln} T$ are calculated and shown as black lines. Although the magnitude of the imaginary part decreases monotonously as temperature decreases,  its logarithmic temperature derivative gives a maximum at a characteristic temperature $T_{\rm coh}$, which corresponds to the coherence of the heavy quasi-particles\cite{Burdin:2001aa,Shim:2007aa,Yang:2008aa}. Therefore, the heavy Fermion liquid coherence happens following the formation of localized Kondo singlet states in the $\gamma$-Ce, while they happen nearly at the same temperature in the heavy Fermion material CeCu$_2$Si$_2$\cite{Reinert:2001aa}. In addition, $T_{\rm coh}$ changes from 40 K for $\gamma$-Ce to 129 K for $\alpha$-Ce.

\section{Discussion}

From our XRD and ARPES experiments, we observed that the $\gamma$-$\alpha$ phase transition occurred at $\approx$ 50 K in the bulk of thin film and the surface remained in the $\gamma$-phase down to 17 K. It means that the $\gamma$ phase in the surface is much more robust than that in the bulk, which can not be solely explained by the strain effect from substrate, due to the fact that the strain effect on the topmost surface is weaker than that in the bulk. Hence, there should be other factors besides strain effect that leads to the different performance of bulk and surface. With the appearance of the Kondo dip in our ARPES results below 80 K, the Kondo effect might be the proper mechanism. 

To clarify the role of Kondo effect, we have performed DFT+DMFT simulation to explore the Kondo hybridization and RKKY interaction, which are the two most important features of Kondo effect\cite{Burdin:2001aa,Shim:2007aa,Yang:2008aa,Zhang:2000aa,Reinert:2001aa}. Firstly, DFT+DMFT calculations confirm that the characteristic temperature of Kondo singlet state formation $T_{\rm KS}$ is about 80 K by observing the low-frequency dip property of $-{\rm Im}[ \Sigma_{5/2} (\omega,T)]$, which agreed well with our ARPES results. Moreover, the characteristic temperature of the heavy fermion liquid coherence due to the RKKY interaction from simulation is about 40 K, which is in the neighbourhood of $T_{\gamma-\alpha}$ (about 50 K) from the XRD spectra. Therefore, the Kondo effect has a close relationship with the $\gamma$-$\alpha$ phase transition. And our work suggests that Kondo scenario might be the mechanism of the phase transition.

Moreover, the Kondo scenario is assumed to work through a positive feedback mechanism together with lattice contraction in the $\gamma$-$\alpha$ phase transition as temperature decreases. Firstly, as temperature decreases, if the lattice constant $a_{\rm lat}$ is fixed at the same value, the strength of effective $f$-$c$ electron hybridization $V_{fc}$ becomes stronger as $T$ decreasing, which enforces $f$-electrons to bond indirectly at the critical temperature $T_{\rm coh}$. Secondly, the $f$-$f$ indirectly bonding would make the ions at different sites attractive, which means that both the energy bands of $f$- and $c$-electrons become wider and their hybridization becomes stronger. Such positive feedback processes would repeat again and again, and induce a first-order isomorphic volume phase transition at last. Besides, as observed in many previous experimental works\cite{James:1952aa,Burgardt:1976aa}, the phase transition temperature from $\alpha$ to $\gamma$ is greater than that from $\gamma$ to $\alpha$, as much stronger thermal fluctuation is demanded to break the $f$-electron coherent bonding in $\alpha$-Ce.

The above-mentioned positive feedback would mean a ``hysteresis'' phenomenon in the transport and XRD experiments for cerium, which has already been studied in literatures for bulk Ce metal \cite{James:1952aa,Burgardt:1976aa,McHargue:1960aa,Gschneidner:1962aa}. As for our 200 nm thin film, we also observed hysteresis in the resistivity between cooling and warming, as shown in Supplementary Figure 5. Although the resistivity curves for bulk Ce metal and Ce thin film differed in the detail, the phenomenon of hysteresis was robust in both bulk Ce metal and Ce thin film, and the ``positive feedback'' scenario might be reasonable.

To summarise, we have experimentally and theoretically investigated the physical mechanism of the $\gamma$-$\alpha$ phase transition in high quality single crystalline cerium thin films. ARPES spectra show the hybridization of $f$ level with conduction bands and the formation of a Kondo dip upon cooling the sample from 300 K to 17 K, and our ARPES results at low temperature indicated the absence of $\alpha$-phase at the sample surface. The varied temperature XRD demonstrated that the $\gamma$-$\alpha$ phase transition occurred around 50 K, which is retarded by strain effect from substrate comparing with $T_{\gamma-\alpha}$ (about 140 K) of the bulk Ce metal. The Kondo effect can be taken advantage to describe these phenomena. From our DFT+DMFT calculations, we obtained the characteristic temperature of Kondo hybridization $T_{\rm KS}$ at about 80 K and the characteristic temperature of heavy fermion liquid coherence $T_{\rm coh}$ at about 40 K. The former one agreed with the ARPES results, and the latter one is in the neighborhood of $T_{\gamma-\alpha}$ from XRD experiments. Therefore, our experimental and theoretical results proposed that the Kondo scenario might play an important role in the $\gamma$-$\alpha$ phase transition  of cerium thin films.

\section{Methods}

\textbf{Sample preparation}. High quality Ce thin films were synthesized by molecular beam epitaxy (MBE). Ce source with 99.95\% purity was used and was thoroughly degassed at 1660 $^\circ$C before evaporation to the graphene substrate grown on 6$H$-SiC(0001). The graphene substrate was prepared following the recipe as described in Ref. \onlinecite{Qi:2010aa,Moreau:2010aa}, with some modification to optimize the quality of the \emph{as-grown} graphene. N-doped 6$H$-SiC(0001) was first annealed in UHV at 600 $^\circ$C for at least 3 hours to remove the absorbed gases or organic molecules. After annealing at $\sim$950 $^\circ$C for 18 minutes under Si flux to produce a (3$\times$3) reconstruction, the substrate was heated to $\sim$ 1400 $^\circ$C for 10 minutes to form graphene.

The Ce source was kept at 1620 $^\circ$C during the evaporation. The base pressure of our MBE system is better than $5.0\times10^{-11}$ mbar and rise to less than $2.0\times10^{-10}$ mbar during Ce evaporation. Even under this ultra high vacuum environment, special care should be taken to avoid the \emph{in-situ} oxidation of the thin films by trace amount residual gases in the UHV chamber during the evaporation process. The Ce source must be degassed at 1660 $^\circ$C for at least 4 hours before each deposition. The thermal radiation from the hot Ce source should be also taken into consideration, as it might heat up the substrate and result in the oxidization of the Ce films by the residual oxygen in the MBE growth chamber. During our MBE growth, the substrate temperature was kept at $\sim$ 80 $^\circ$C. The flux rate of Ce was determined by Quartz crystal micro-balance (QCM) and was $\sim$ 0.127 $\text{\AA}$/S at 1620 $^\circ$C.

\textbf{VT-XRD experiments}. For the varied temperature XRD experiment, the Ce thin films were packed in a glove box under Argon atmosphere after taken out from the MBE chamber, and then transferred to a Panalytical Empyrean X-Ray Diffractometer. The sample stage of the diffractometer could be cooled by a close-cycled Helium refrigerator and heated by the heating stage. The sample temperature could be set continiously from room temperature to 15 K. We have taken the XRD spectra on a 200 nm thick Ce thin film from 300 K to 20 K with 20 K intervals, and also at the lowest achievable temperature 15 K.

\textbf{ARPES experiments}. For the photoemission spectroscopy experiment, the spectra were excited by the He I$\alpha$ (21.2 eV) resonance line of a commercial Helium gas discharge lamp. The light was guided to the analysis chamber by a quartz capillary. In virtue of the efficient four-stage differential pumping system, the pressure in the analysis chamber was better than $1.0\times10^{-10}$ mbar during our experiments, satisfying the harsh criteria for probing the physical properties of chemically high reactive lanthanide/actinide elements. A VG Scienta R4000 energy analyzer was used to collect the photoelectrons.

\textbf{HR-STEM experiments}. The HAADF-STEM experiment was performed in a FEI Titan G2 80-200 at 300 kV, which was equipped with an aberration corrector and provided us a sufficiently high spatial resolution ($\sim$ 0.7 \AA). The convergence angle of the probe was $\sim$ 21.4 mrad and the screen beam current was about 0.5 nA. Before the experiment, all of the low-order aberrations have been adjusted to an acceptable level. For high-resolution STEM imaging, the size of the HRSTEM micrography was 1024 $\times$ 1024 pixels, with a dwell time of 8 $\mu$s.

\textbf{Theoretical calculations}. To explore the properties of electronic structure and microscopy mechanism of the phase transition of Ce, we performed the theoretical calculations combining the density functional theory and the dynamical mean-filed theory method (DFT+DMFT) based on the eDMFT software package \cite{Haule2010}. We used WIEN2k package based on the full-potential linearized augmented plane-wave (LAPW) method for the density function theory part \cite{Blaha:2001aa,Blaha2010}. The lattice parameters were taken from our XRD experiments. Perdew-Burke-Ernzerhof generalized gradient approximation (GGA) was used for the exchange-correlation functional with 5000 k-point meshes for the whole Brillouin zone and as the Ce-ion is so heavy,  the spin-orbital coupling (SOC) effect was considered. During all the calculations, we set the Muffin-tin radii to 2.50 a.u., R$_{\rm MT}$K$_{\rm MAX}$ = 8.0 and G$_{\rm MAX}$ = 14.0 .  For the DMFT part, the one-crossing approximation (OCA) was used as the impurity solver as the DMFT method mapping  the derived lattice model to a single-impurity model \cite{Haule2001,Shim2007a,Haule2010}. The Coulomb interaction $U$ on the Ce f-orbital was set to 6.0 eV with Hund interaction $J$ to 0.7 eV.

\textbf{Acknowledgments}

We thank Zheng Liu, Jia-Wei Mei, Yi-Feng Yang, Yuan-Ji Xu, Guang-Ming Zhang, Hai-Yan Lu, Li Huang and Yin Zhong for helpful discussions. The work was supported by the Science Challenge Project (NO.TZ2016004 and NO.TZ2018002), the Foundation of President of CAEP (NO.201501040), the National Basic Research Program of China (NO.2015CB921303),  the National Key Research and Development Program of China (NO.2017YFA0303104), the National Science Foundation of China (NO.U1630248), the Open Research Fund Program of the State Key Laboratory of Low-Dimensional Quantum Physics (No.KF201702) and the SPC-Lab Research Fund (NO.XKFZ201605 and NO.WDZC201901).
We thank  the Tianhe-1A platforms at the National Supercomputer Center in Tianjin.

\textbf{Author contributions}

X. G. Zhu, Yu Liu and Y. W. Zhao contributed equally to this work. X.G., Y.Z., Y.D., W.F., Y.H., W.Z., D.H. and Yi.L. carried out the growth of thin films and XRD experiments. Y.W. and C.L. performed the HR-STEM experiments. Yu.L., Y.C., D.J. and X.B. did the theoretical calculations. X.G., S.Y., Q.L., L.Z. and Q.Y. performed the ARPES measurements. X.G., H.F. and X.C. conceived the experiments and wrote the manuscript.

\textbf{Competing financial interests:} The Authors declare no Competing Financial or Non-Financial Interests.

\textbf{Data Availability:} The datasets generated during and/or analysed during the current study are available from the corresponding author on reasonable request.

\clearpage

\begin{thebibliography}{10}
\expandafter\ifx\csname url\endcsname\relax
  \def\url#1{\texttt{#1}}\fi
\expandafter\ifx\csname urlprefix\endcsname\relax\def\urlprefix{URL }\fi
\expandafter\ifx\csname href\endcsname\relax
  \def\href#1#2{#2} \def\path#1{#1}\fi

\bibitem{Alexander:2012aa}
Nikolaev, A. V. \& Tsvyashchenko, A. V.
  \href{http://stacks.iop.org/1063-7869/55/i=7/a=R02}{The puzzle of the $\gamma \rightarrow \alpha$ and other phase transitions in cerium}. Phys-Usp. \textbf{55}, 657-680 (2012).

\bibitem{Tsiok:2001aa}
Tsiok, O. B. \& Khvostantsev, L. G. \href{https://doi.org/10.1134/1.1435746}{Phase transitions in cerium at high pressures (up to 15 GPa) and high temperatures}. J. Exp. Theor. Phys. \textbf{93}, 1245-1249 (2001).

\bibitem{Schiwek:2002aa}
Schiwek, A. Porsch, F. \& Holzapfel, W. B.
  \href{https://doi.org/10.1080/08957950212799}{High temperature-high pressure structural studies of cerium}. High Pressure Res. \textbf{22}, 407-410 (2002).

\bibitem{Lipp:2008aa}
Lipp, M. J., et al.
  \href{https://link.aps.org/doi/10.1103/PhysRevLett.101.165703}{Thermal signatures of the Kondo volume collapse in cerium}. Phys. Rev. Lett. \textbf{101}, 165703 (2008).

\bibitem{Tian:2015aa}
Tian, M. F., et al.
  \href{https://link.aps.org/doi/10.1103/PhysRevB.91.125148}{Thermodynamics of the $\alpha$-$\gamma$ transition in cerium studied by an LDA + Gutzwiller method}. Phys. Rev. B \textbf{91}, 125148 (2015).

\bibitem{James:1952aa}
James, N. R., Legvold, S. \& Spedding, F. H.
  \href{http://link.aps.org/doi/10.1103/PhysRev.88.1092}{The resistivity of
  lanthanum, cerium, praseodymium, and neodymium at low temperatures}. Phys. Rev. \textbf{88}, 1092-1098 (1952).

\bibitem{Burgardt:1976aa}
Burgardt, P., et al.
  \href{http://link.aps.org/doi/10.1103/PhysRevB.14.2995}{Electrical resistivity and magnetic susceptibility of $\beta$-cerium from 2 to 300 K}. Phys. Rev. B \textbf{14}, 2995-3006 (1976).

\bibitem{Wilkinson:1961aa}
Wilkinson, M. K., Child, H. R., McHargue, C. J., Koehler, W. C. \& Wollan, E. O.
  \href{http://link.aps.org/doi/10.1103/PhysRev.122.1409}{Neutron diffraction investigations of metallic cerium at low temperatures}. Phys. Rev. \textbf{122}, 1409-1413 (1961).

\bibitem{Coqblin:1968aa}
Coqblin, B. \& Blandin, A.
  \href{https://doi.org/10.1080/00018736800101306}{Stabilit{\'e}des moments magn{\'e}tiques localis{\'e}s dans les m{\'e}taux}. Adv. Phys. \textbf{17}, 281-366 (1968).

\bibitem{Ramirez:1971aa}
Ramirez, R. \& Falicov, L. M.
  \href{https://link.aps.org/doi/10.1103/PhysRevB.3.2425}{Theory of the $\alpha$-$\gamma$ phase transition in metallic cerium}. Phys. Rev. B \textbf{3}, 2425-2430 (1971).

\bibitem{Hirst:1974aa}
Hirst, L. L.
  \href{http://www.sciencedirect.com/science/article/pii/S0022369774801538}{Configuration crossover in 4$f$ substances under pressure}. J. Phys. Chem. Solids \textbf{35}, 1285-1296 (1974).

\bibitem{Gustafson:1969aa}
Gustafson, D. R., McNutt, J. D. \& Roellig, L. O.
  \href{https://link.aps.org/doi/10.1103/PhysRev.183.435}{Positron annihilation in $\gamma$- and $\alpha$- cerium}. Phys. Rev. \textbf{183}, 435-440 (1969).

\bibitem{Gempel:1972aa}
Gempel, R. F., Gustafson, D. R. \& Willenberg, J. D.
  \href{https://link.aps.org/doi/10.1103/PhysRevB.5.2082}{Investigation of the electronic structure of $\alpha^{\prime}$-cerium by angular-correlation studies of positron-annihilation radiation}. Phys. Rev. B \textbf{5}, 2082-2085 (1972).

\bibitem{Kornstadt:1980aa}
Kornst\"adt, U., L\"asser, R. \& Lengeler, B.
  \href{https://link.aps.org/doi/10.1103/PhysRevB.21.1898}{Investigation of the $\gamma$-$\alpha$ phase transition in cerium by Compton scattering}. Phys. Rev. B \textbf{21}, 1898-1901 (1980).

\bibitem{Murani:2002aa}
Murani, A. P., Reske, J., Ivanov, A. S. \& Palleau, P.
  \href{https://link.aps.org/doi/10.1103/PhysRevB.65.094416}{Evolution of the spin-orbit excitation across the $\gamma \rightarrow \alpha$ transition in Ce}. Phys. Rev. B \textbf{65}, 094416 (2002).

\bibitem{Murani:2005aa}
Murani, A. P., Levett, S. J. \& Taylor, J. W.
  \href{https://link.aps.org/doi/10.1103/PhysRevLett.95.256403}{Magnetic form factor of $\alpha$-Ce: Towards understanding the magnetism of cerium}. Phys. Rev. Lett. \textbf{95}, 256403 (2005).

\bibitem{Johansson:1974ab}
Johansson, B.
  \href{http://dx.doi.org/10.1080/14786439808206574}{The $\alpha$-$\gamma$ transition in cerium is a Mott transition}. Philos. Mag. \textbf{30}, 469-482 (1974).

\bibitem{Allen:1982aa}
Allen, J. W. \& Martin, R. M.
  \href{http://link.aps.org/doi/10.1103/PhysRevLett.49.1106}{Kondo volume collapse and the $\gamma \rightarrow \alpha$ transition in cerium}. Phys. Rev. Lett. \textbf{49}, 1106-1110 (1982).

\bibitem{Allen:1992aa}
Allen, J. W. \& Liu, L. Z.
  \href{https://link.aps.org/doi/10.1103/PhysRevB.46.5047}{$\alpha$-$\gamma$ transition in Ce. ii. a detailed analysis of the Kondo volume-collapse model}. Phys. Rev. B \textbf{46}, 5047-5054 (1992).

\bibitem{Zhang:2000aa}
Zhang, G. M. \& Yu, L.
  \href{http://link.aps.org/doi/10.1103/PhysRevB.62.76}{Kondo singlet state coexisting with antiferromagnetic long-range order: A possible ground state for Kondo insulators}. Phys. Rev. B \textbf{62}, 76-79 (2000).

\bibitem{Burdin:2001aa}
Burdin, S., Georges, A. \& Grempel, D. R.
  \href{https://link.aps.org/doi/10.1103/PhysRevLett.85.1048}{Coherence scale of the Kondo lattice}. Phys. Rev. Lett. \textbf{85}, 1048-1051 (2000).

\bibitem{Reinert:2001aa}
Reinert, F., et al.
  \href{http://link.aps.org/doi/10.1103/PhysRevLett.87.106401}{Temperature dependence of the Kondo resonance and its satellites in CeCu$_2$Si$_2$}. Phys. Rev. Lett. \textbf{87}, 106401 (2001).

\bibitem{Yang:2008aa}
Yang, Y. F., Fisk, Z., Lee, H. O., Thompson, J. D. \& Pines, D.
  \href{https://doi.org/10.1038/nature07157}{Scaling the Kondo lattice}. Nature \textbf{454}, 611-613 (2008).

\bibitem{Stassis:1979aa}
Stassis, C., Gould, T., McMasters, O. D., Gschneidner, K. A. \& Nicklow, R. M.
  \href{http://link.aps.org/doi/10.1103/PhysRevB.19.5746}{Lattice and spin dynamics of $\gamma$-Ce}. Phys. Rev. B \textbf{19}, 5746-5753 (1979).

\bibitem{Stassis:1982aa}
Stassis, C., Loong, C. K., McMasters, O. D. \& Nicklow, R. M.
  \href{https://link.aps.org/doi/10.1103/PhysRevB.25.6485}{Addendum to the lattice dynamics of $\gamma$-Ce}. Phys. Rev. B \textbf{25}, 6485-6487 (1982).

\bibitem{Krisch:2011aa}
Krisch, M., et al.
  \href{http://www.ncbi.nlm.nih.gov/pmc/articles/PMC3111256/}{Phonons of the anomalous element cerium}. P. Natl. Acad. Sci. USA \textbf{108}, 9342-9345 (2011).

\bibitem{Weschke:1998aa}
Weschke, E., et al.
  \href{http://link.aps.org/doi/10.1103/PhysRevB.58.3682}{Surface electronic structure of epitaxial Ce and La films}. Phys. Rev. B \textbf{58}, 3682-3689 (1998).

\bibitem{Schiller:2003aa}
Schiller, F., Heber, M., Servedio, V. D. P. \& Laubschat, C.
  \href{http://link.aps.org/doi/10.1103/PhysRevB.68.233103}{Surface states and Fermi surface of ordered $\gamma$-like Ce films on W(110)}. Phys. Rev. B \textbf{68}, 233103 (2003).

\bibitem{Victor:2015aa}
Alexandrov, V., Coleman, P. \& Erten, O.
  \href{https://link.aps.org/doi/10.1103/PhysRevLett.114.177202}{Kondo breakdown in topological Kondo insulators}. Phys. Rev. Lett. \textbf{114}, 177202 (2015).

\bibitem{Erten:2016aa}
Erten, O., Ghaemi, P. \& Coleman, P.
  \href{https://link.aps.org/doi/10.1103/PhysRevLett.116.046403}{Kondo breakdown and quantum oscillations in SmB$_6$}. Phys. Rev. Lett. \textbf{116}, 046403 (2016).

\bibitem{Robert:2016aa}
Peters, R., Yoshida, T., Sakakibara, H. \& Kawakami, N.
  \href{https://link.aps.org/doi/10.1103/PhysRevB.93.235159}{Coexistence of light and heavy surface states in a topological multiband Kondo insulator}. Phys. Rev. B \textbf{93}, 235159 (2016).

\bibitem{Chen:2018aa}
Chen, Q. Y., et al.
  \href{https://doi.org/10.1103/PhysRevB.97.155155}{Localized to itinerant transition of $f$ electrons in ordered Ce films on W(110)}. Phys. Rev. B \textbf{97}, 155155 (2018).

\bibitem{Morimoto:2009aa}
Morimoto, O., Kato, H., Enta, Y. \& Sakisaka, Y.
  \href{http://www.sciencedirect.com/science/article/pii/S0039602809003021}{Photoemission from the valence bands of Ce(111) on W(110)}. Surf. Sci. \textbf{603}, 2145-2151 (2009).

\bibitem{Koskenmaki:1978ab}
Koskenmaki, D. C. \& Gschneidner, K. A.
  \href{https://doi.org/10.1016/S0168-1273(78)01008-9}{\emph{Handbook on the Physics and Chemistry of Rare Earths, Chapter 4: Cerium.} (Elsevier Press, 1978)}.

\bibitem{Haule2010}
Haule, K., Yee, C. H. \& Kim, K.
  \href{https://link.aps.org/doi/10.1103/PhysRevB.81.195107}{Dynamical mean-field theory within the full-potential methods: Electronic structure of CeIrIn$_5$, CeCoIn$_5$ and CeRhIn$_5$}. Phys. Rev. B \textbf{81}, 195107 (2010).

\bibitem{Blaha:2001aa}
Blaha, P., Schwarz, K., Madsen, G., Kvasnicka, D. \& Luitz, J.
  Wien2k: An augmented plane wave plus local orbitals program for calculating crystal properties. Techn. Universitat Wien, 2008.

\bibitem{Qi:2010aa}
Qi, Y., Rhim, S. H., Sun, G. F., Weinert, M. \& Li, L.
  \href{http://link.aps.org/doi/10.1103/PhysRevLett.105.085502}{Epitaxial graphene on SiC(0001): More than just honeycombs}. Phys. Rev. Lett. \textbf{105}, 085502 (2010).

\bibitem{Moreau:2010aa}
Moreau, E., et al.
  \href{http://scitation.aip.org/content/aip/journal/apl/97/24/10.1063/1.3526720}{Graphene growth by molecular beam epitaxy on the carbon-face of SiC}. Appl. Phys. Lett. \textbf{97}, 241907 (2010).

\bibitem{Blaha2010}
Blaha, P., Schwarz, K., Madsen, G., Kvasnicka, D. \& Luitz, J.
  Wien2k: An augmented plane wave plus local orbitals program for calculating crystal properties. J. Endocrinol. \textbf{196}, 123-130 (2010).

\bibitem{Haule2001}
Haule, K., Kirchner, S., Kroha, J. \& W\"olfle, P.
  \href{https://link.aps.org/doi/10.1103/PhysRevB.64.155111}{Anderson impurity model at finite Coulomb interaction $U$: Generalized noncrossing approximation}. Phys. Rev. B \textbf{64}, 155111 (2001).

\bibitem{Shim2007a}
Shim, J. H., Haule, K. \& Kotliar, G.
  \href{https://doi.org/10.1038/nature05647}{Fluctuating valence in a correlated solid and the anomalous properties of $\delta$-plutonium}. Nature \textbf{446}, 513-516 (2007).

\bibitem{Bauer:1998aa}
Bauer, A., et al.
  \href{https://link.aps.org/doi/10.1103/PhysRevB.57.2647}{High-precision determination of atomic positions in crystals: The case of 6$H$- and 4$H$-SiC}. Phys. Rev. B \textbf{57}, 2647-2650 (1998).

\bibitem{Patthey:1985aa}
Patthey, F., Delley, B., Schneider, W. D. \& Baer, Y.
  \href{http://link.aps.org/doi/10.1103/PhysRevLett.55.1518}{Low-energy excitations in $\alpha$- and $\gamma$-Ce observed by photoemission}. Phys. Rev. Lett. \textbf{55}, 1518-1521 (1985).

\bibitem{Weschke:1991aa}
Weschke, E., et al.
  \href{http://link.aps.org/doi/10.1103/PhysRevB.44.8304}{Surface and bulk electronic structure of Ce metal studied by high-resolution resonant photoemission}. Phys. Rev. B \textbf{44} 8304-8307 (1991).

\bibitem{Kucherenko:2002aa}
Kucherenko, Y., Molodtsov, S. L., Heber, M. \& Laubschat, C.
  \href{https://link.aps.org/doi/10.1103/PhysRevB.66.155116}{4$f$-derived electronic structure at the surface and in the bulk of $\alpha$-Ce metal}. Phys. Rev. B \textbf{66}, 155116 (2002).

\bibitem{Chen:2017aa}
Chen, Q. Y., et al.
  \href{https://link.aps.org/doi/10.1103/PhysRevB.96.045107}{Direct observation of how the heavy-fermion state develops in CeCoIn$_5$}. Phys. Rev. B \textbf{96}, 045107 (2017).

\bibitem{Gu:1991ab}
Gu, C., Wu, X., Olson, C. G. \& Lynch, D. W.
  \href{http://dx.doi.org/10.1103/PhysRevLett.67.1622}{``$\gamma$-$\alpha$'' phase transition of monolayer Ce on W(110)}. Phys. Rev. Lett. \textbf{67}, 1622-1625 (1991).

\bibitem{Vyalikh:2006aa}
Vyalikh, D. V., et al.
  \href{http://dx.doi.org/10.1103/PhysRevLett.96.026404}{Wave-vector conservation upon hybridization of 4$f$ and valence-band states observed in photoemission spectra of a Ce monolayer on W(110)}. Phys. Rev. Lett. \textbf{96}, 026404 (2006).

\bibitem{Millhouse:1974aa}
Millhouse, A. H. \& Furrer, A.
  \href{http://www.sciencedirect.com/science/article/pii/0038109874913672}{Crystal field splitting in $\gamma$-cerium}. Solid Stat. Commun. \textbf{15}, 1303-1306 (1974).

\bibitem{Shim:2007aa}
Shim, J. H., Haule, K. \& Kotliar, G.
  \href{http://science.sciencemag.org/content/318/5856/1615.abstract}{Modeling the localized-to-itinerant electronic transition in the heavy fermion system CeIrIn$_5$}. Science \textbf{318}, 1615-1617 (2007).

\bibitem{McHargue:1960aa}
McHargue, G. J. \& Yakel, H. L.
  \href{http://www.sciencedirect.com/science/article/pii/0001616060900195}{Phase transformations in cerium}. Acta Metallurgica \textbf{8}, 637-646 (1960).

\bibitem{Gschneidner:1962aa}
Gschneidner, K. A., Elliott, R. O. \& McDonald, R. R.
  \href{http://www.sciencedirect.com/science/article/pii/0022369762905139}{Effects of alloying additions on the $\gamma\rightleftharpoons\alpha$ transformation of cerium - Part I. Pure cerium}. J. Phys. Chem. Solids \textbf{23}, 555-566 (1962).

\end{thebibliography}
\end{document}


\captionsetup[figure]{labelfont={bf},labelformat={default},labelsep=period,name={Supplementary Figure }}

\title{Supporting Online Materials: Kondo scenario of the $\gamma$-$\alpha$ phase transition in single crystalline Cerium thin films}

\author{Xie-Gang Zhu$^\star$}
\affiliation{Science and Technology on Surface Physics and Chemistry Laboratory, Jiangyou 621908, Sichuan, China}

\author{Yu Liu$^\star$}
\affiliation{Laboratory of Computational Physics, Institute of Applied Physics and Computational Mathematics, Beijing 100088, China}
\affiliation{Software Center for High Performance Numerical Simulation,
China Academy of Engineering Physics, Beijing 100088, China}

\author{Ya-Wen Zhao$^\star$}
\affiliation{Institute of Materials, China Academy of Engineering Physics, Jiangyou 621908, Sichuan, China}

\author{Yue-Chao Wang}
\affiliation{Laboratory of Computational Physics, Institute of Applied Physics and Computational Mathematics, Beijing 100088, China}

\author{Yun Zhang}
\affiliation{Science and Technology on Surface Physics and Chemistry Laboratory, Jiangyou 621908, Sichuan, China}

\author{Chao Lu}
\affiliation{Institute of Materials, China Academy of Engineering Physics, Jiangyou 621908, Sichuan, China}

\author{Yu Duan}
\affiliation{Science and Technology on Surface Physics and Chemistry Laboratory, Jiangyou 621908, Sichuan, China}

\author{Dong-Hua Xie}
\affiliation{Institute of Materials, China Academy of Engineering Physics, Jiangyou 621908, Sichuan, China}

\author{Wei Feng}
\affiliation{Science and Technology on Surface Physics and Chemistry Laboratory, Jiangyou 621908, Sichuan, China}

\author{Dan Jian}
\affiliation{Science and Technology on Surface Physics and Chemistry Laboratory, Jiangyou 621908, Sichuan, China}
\affiliation{Laboratory of Computational Physics, Institute of Applied Physics and Computational Mathematics, Beijing 100088, China}

\author{Yong-Huan Wang}
\affiliation{Science and Technology on Surface Physics and Chemistry Laboratory, Jiangyou 621908, Sichuan, China}

\author{Shi-Yong Tan}
\affiliation{Science and Technology on Surface Physics and Chemistry Laboratory, Jiangyou 621908, Sichuan, China}

\author{Qin Liu}
\affiliation{Science and Technology on Surface Physics and Chemistry Laboratory, Jiangyou 621908, Sichuan, China}

\author{Wen Zhang}
\affiliation{Science and Technology on Surface Physics and Chemistry Laboratory, Jiangyou 621908, Sichuan, China}
\affiliation{Sate Key Laboratory of Environmental-Friendly Energy Materials, Southwest University of Science and Technology, Mianyang 621010, Sichuan, China}

\author{Yi Liu}
\affiliation{Science and Technology on Surface Physics and Chemistry Laboratory, Jiangyou 621908, Sichuan, China}

\author{Li-Zhu Luo}
\affiliation{Science and Technology on Surface Physics and Chemistry Laboratory, Jiangyou 621908, Sichuan, China}

\author{Xue-Bing Luo}
\affiliation{Science and Technology on Surface Physics and Chemistry Laboratory, Jiangyou 621908, Sichuan, China}

\author{Qiu-Yun Chen}
\email{sheqiuyun@126.com}
\affiliation{Science and Technology on Surface Physics and Chemistry Laboratory, Jiangyou 621908, Sichuan, China}

\author{Hai-Feng Song}
\email{song\_haifeng@iapcm.ac.cn}
\affiliation{Laboratory of Computational Physics, Institute of Applied Physics and Computational Mathematics, Beijing 100088, China}
\affiliation{Software Center for High Performance Numerical Simulation,
China Academy of Engineering Physics,
Beijing 100088, China}

\author{Xin-Chun Lai}
\email{laixinchun@caep.cn}
\affiliation{Science and Technology on Surface Physics and Chemistry Laboratory, Jiangyou 621908, Sichuan, China}

\date{\today}

\maketitle

\textbf{1. Lattice match between Graphene/SiC substrate and \emph{as-grown} Ce thin film}

We schematically demonstrate the stacking sequence of Ce atoms on Graphene substrate grown on SiC(0001) in Supplementary Figure \ref{lat}. The in-plane lattice parameter of SiC(0001) surface is 3.08 \AA\ (Supplementary Ref. \cite{Bauer:1998aa}). And the \emph{as-grown} Graphene forms a $6\sqrt{3}\times6\sqrt{3}R30^\circ$ reconstructed structure on SiC(0001) surface (Supplementary Ref. \cite{Huang:2008aa}), as shown by the outer black rhombus in Supplementary Figure \ref{lat}. The in-plane lattice parameter of $\gamma$-phase Ce is calculated to be 3.65 \AA\ (Supplementary Ref. \cite{Alexander:2012aa}). We noticed that Ce atoms could be seated in a \emph{pseudo}-6$\times$6 supercell of SiC(0001) surface, as indicated by the red rhombus in Supplementary Figure \ref{lat}. The word \emph{pseudo} was used here in the sense that the center of the yellow hexagon in Supplementary Figure \ref{lat} matches no lattice point of SiC(0001) surface or Graphene. What is more, Graphene is stable and inert, while Ce is highly chemically reactive. Therefore, we adopted Graphene grown on SiC(0001) as the substrate for the epitaxial growth of Ce thin films.

\begin{figure*}[h]
\centering
\includegraphics[width=1.0\textwidth]{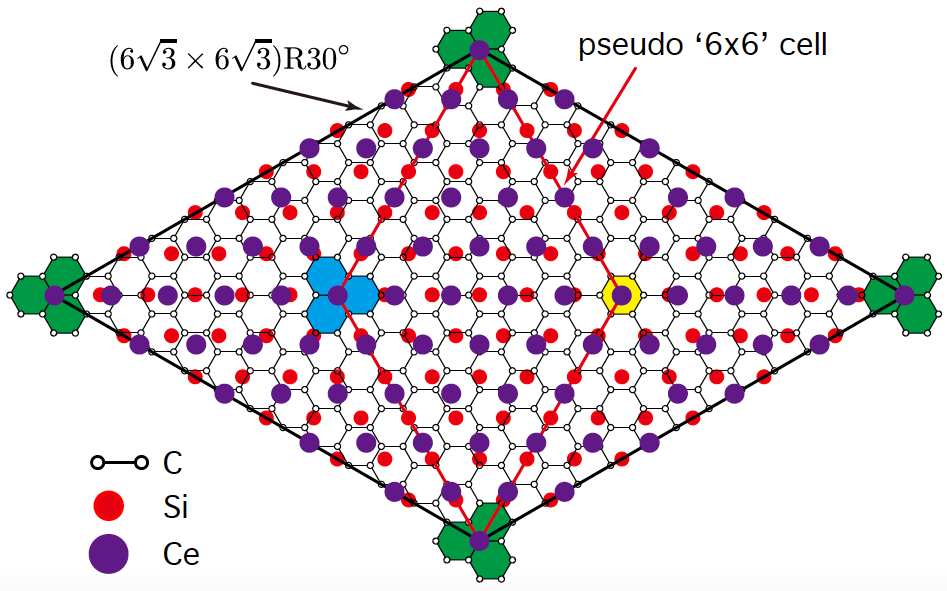}
\caption{Schematical stacking sequence of Ce atoms on Graphene substrate grown on SiC(0001).}
\label{lat}
\end{figure*}

\textbf{2. Thickness dependant ARPES}

The ARPES results of a 200 nm and a 40 nm thick Ce thin films at 300 K, 80 K and 17 K were shown in Supplementary Figure \ref{varT}. The valence band of 200 nm at 300 K in Supplementary Figure \ref{varT} was fitted by a parabolic curve and the dispersion at 80 K was fitted by the PAM model function as mentioned in the manuscript. We could see that the dispersions of the samples for 200 nm at 80 K and 40 nm at 17 K behaved similar, \emph{i.e.,}, they both showd strong hybridization effect between the valence bands and $4f^1$ level.

\begin{figure*}[h]
\centering
\includegraphics[width=1.0\textwidth]{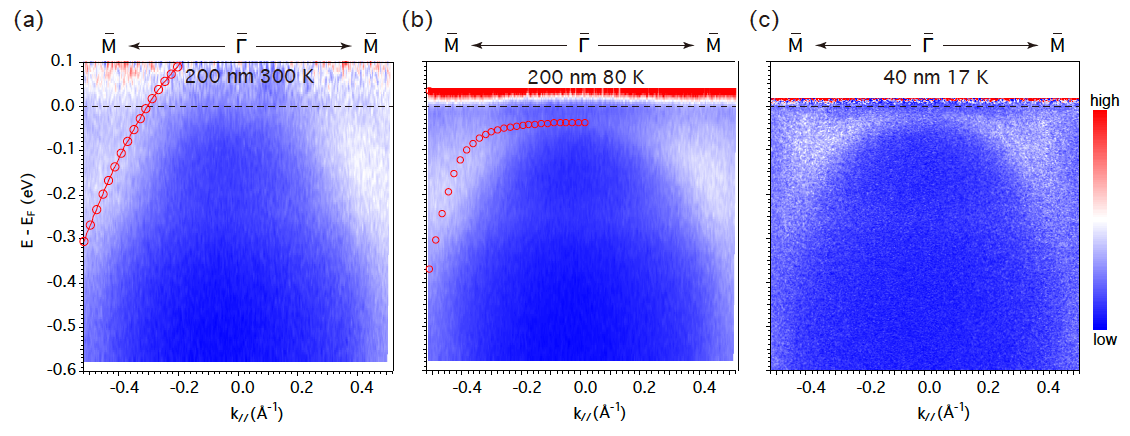}
\caption{(a) and (b) ARPES spectra of a 200 nm Ce film at 300 K and 80 K. (c) ARPES spectrum of a 40 nm Ce film at 17 K.}
\label{varT}
\end{figure*}

\begin{figure*}[h]
\centering
\includegraphics[width=1.0\textwidth]{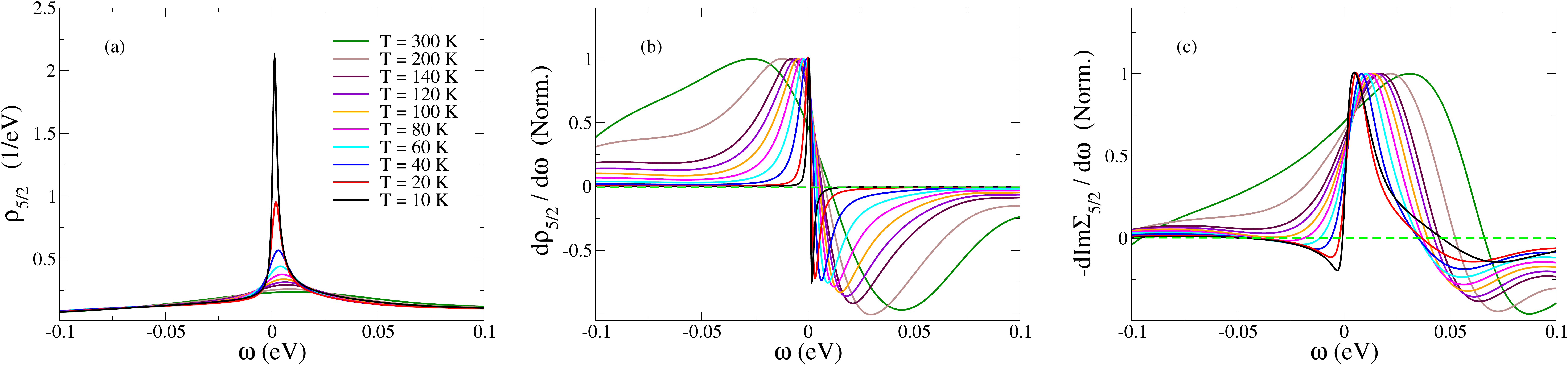}
\caption{The DFT+DMFT results associated with density of states of $\gamma$-Ce  at different temperatures: (a) The partial density of states of 4$f$ ($J$=5/2); (b) The derivative of $4f_{5/2}$'s partial 	density of states with respect to energy $\omega$; (c) The derivative of imaginary part of $4f_{5/2}$'s  self-energy with respect to energy $\omega$. The magnitude of derivatives in (b) and (c) are normalized by divided with the maximum of the derivative at each temperature respectively.}
\label{DOS}
\end{figure*}

\textbf{3. Partial Density of States (PDOS) and Self Energy}

The DFT+DMFT results of $4f_{5/2}$'s partial density of states (PDOS) and the derivatives of PDOS and imaginary part of self-energy (-Im$\Sigma$) with respect to energy $\omega$ at different temperatures are shown in Supplementary Figure \ref{DOS}. For the $4f$ ($J=5/2$) dominates the DOS near Fermi energy, only results of $4f$ ($J=5/2$) are displayed in Supplementary Figure \ref{DOS} (a). It can be clearly seen from Supplementary Figure \ref{DOS}(c) that the derivative of imaginary part of self-energy starts to cross the $y=0$ line just near the Fermi energy at a temperature lower than 80 K. This behavior means that an extrema in self-energy appears below 80 K, which is consistent with the appearance of Kondo-dip near $E_F$ as cooling down to 80 K in our experimental results. However, as only certain particular Feynman diagrams are involved in the OCA procedure, the double-peak structure of Kondo-dip is missing in our simulation. And a double-peak Kondo-dip might be expected from a more accurate and delicate simulation that is based on continuous-time quantum Monte Carlo (CT-QMC) impurity solver.

\textbf{4. NES and AIPES}

NES (Normal Emission Spectrascopy) and AIPES (Angle-Integrated Photoemission Spectroscopy) data were extracted from ARPES spectra with large energy range (-3.0 eV to 0.5 eV in binding energy) for Ce thin films at 300 K, 80 K and 17 K, which correspond to Fig. 4 in the manuscript. As shown in Supplementary Figure \ref{vtPE}(a) and (b), both the NES and AIPES at all three temperatures demonstrate the $4f^0$ ionization peak within a range of -1.90 $\pm$ 0.03 eV, which matches very well with that of $\gamma$-phase Ce in Supplementary Ref. \cite{Weschke:1991aa} ($\approx$ -2.0 eV) and \cite{Kucherenko:2002aa} ($\approx$ -1.9 eV), while they gave $4f^0$ peak positions at -2.2 eV and -2.1 eV for $\alpha$-phase Ce, respectively.

Additionally, we also did data analysis (Lorentzian peak fitting) of the AIPES spectra in Fig. 1 of Supplementary Ref. \cite{Patthey:1985aa} and got the $4f^0$ peak positions as follows: (1) $\alpha$-Ce at 10 K, single peak at -2.37 eV; (2) $\gamma$-Ce at 150 K, double peaks at -2.20 eV and -1.91 eV, respectively. The existence of the double-peak structures for $\gamma$-Ce at 150 K indicates that the sample for PES at 150 K in the literature was most probably a mixture of both $\gamma$- and $\alpha$-phase. The peak at -1.93 eV for the $\gamma$-Ce matches our results.

\begin{figure*}[h]
\centering
\includegraphics[width=1.0\textwidth]{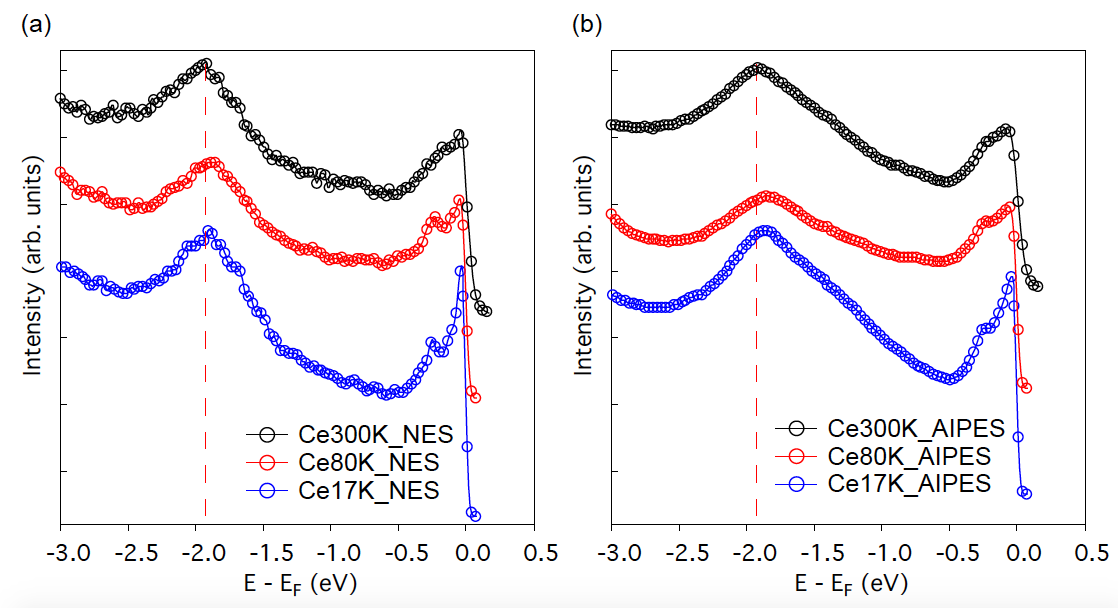}
\caption{(a) and (b) NES and AIPES for Ce thin films at 17 K, 80 K and 300 K, respectively.}
\label{vtPE}
\end{figure*}

\textbf{5. LEED experiments}

A 20 nm thick Ce films was transferred \emph{in-situ} from the MBE chamber to the ARPES chamber within 10 min after the growth. The Low Energy Electron Diffraction (LEED) was taken at 80 K in the ARPES chamber. The \emph{in-plane} lattice parameter of Ce at 80 K was calculated by calibrating with the LEED pattern of a Si(111)-$7\times7$ reconstructed surface at 80 K. The calculated value is 3.60 $\pm$ 0.02 \AA\  and is comparable to the value calculated from the VT-XRD results, \emph{i.e.,}, $\approx$ 3.62 \AA. Therefore, the surface of Ce film remains in the $\gamma$-phase upon cooling to 80 K. We did not take the LEED data at lower temperature, \emph{e.g.}, 17 K (the lowest ARPES experiment temperature). However, by combing the PAM fitting procedure in Fig. 4 in the main context, we could argure that the surface of the Ce thin film remains in the $\gamma$-phase upon cooling to 17 K. The LEED patterns of Ce and Si(111)-7$\times$7 at 80 K are shown in Supplementary Figure \ref{leed}.

\begin{figure*}[h]
\centering
\includegraphics[width=0.75\textwidth]{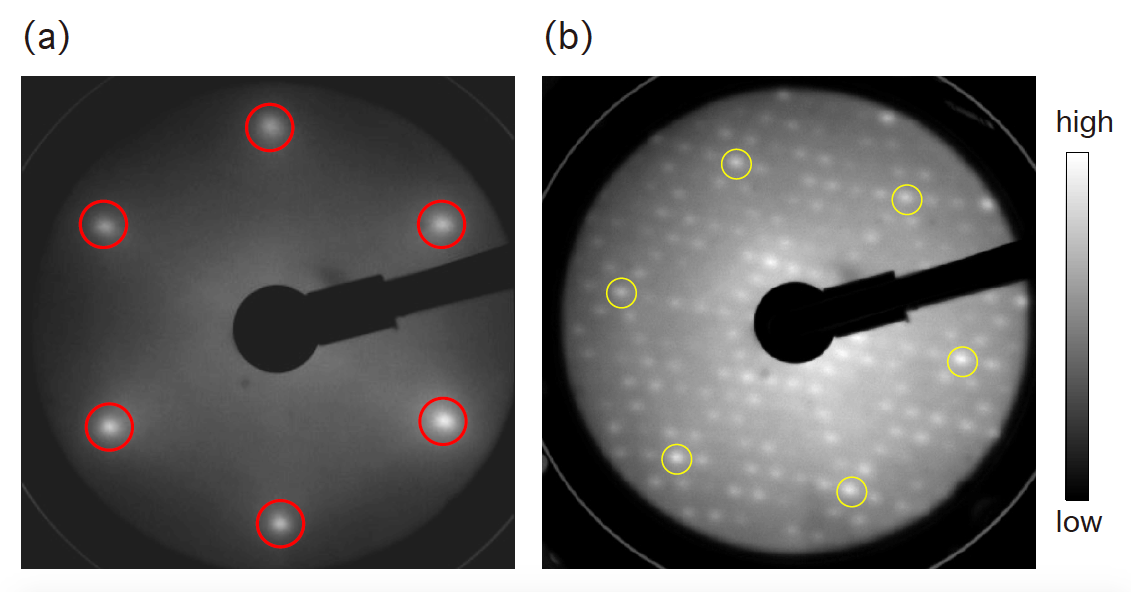}
\caption{(a) and (b) LEED patterns of Ce thin films and Si(111)-$7\times7$ surface at 80 K, respectively. The red and yellow circles indicate the un-reconstructed 1$\times$1 diffraction pattern of Ce and Si surfaces, respectively.}
\label{leed}
\end{figure*}

\textbf{6. Resistivity measurements}

For transport measurements, semi-insulating SiC substrate was used to eliminate the transport contribution from the substrate. The transport measurements were performed with the standard four-probe technique using a physical property measurement system (PPMS-9). Pure Indium (5N) metal was used to make the electrode contacts. The resistivity versus temperature curves of a 200 nm Ce thin film were recorded during a cooling (from 300 K to 2 K) and warming cycle (from 2 K to 300 K), as shown in Supplementary Figure \ref{res}. The cooling and warming curves demonstrated clearly hysteresis behavior, which was consistent with the transport and XRD measurements for bulk Ce metals in literatures (Supplementary Ref. \cite{James:1952aa,McHargue:1960aa,Gschneidner:1962aa,Burgardt:1976aa}). It should be noted that the exact values of the $\gamma$-$\alpha$ and $\alpha$-$\gamma$ phase transition temperatures could not be precisely determined from the transport measurements, due to the fact that the phase transitions were not complete in our sample.

\begin{figure*}[h]
\centering
\includegraphics[width=0.6\textwidth]{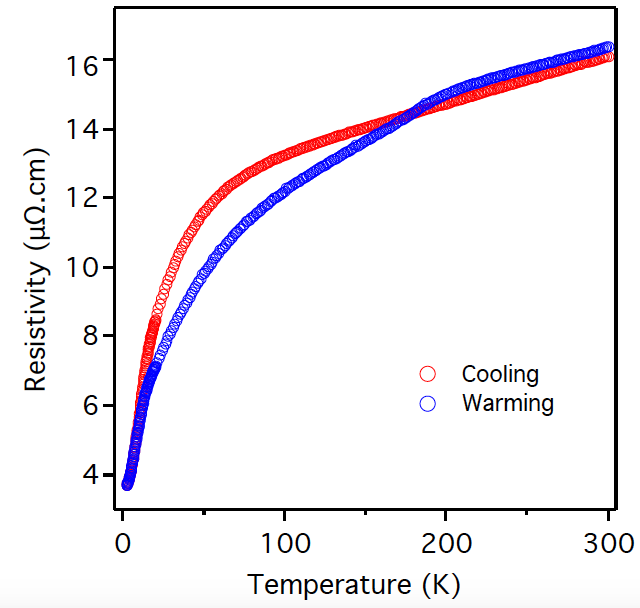}
\caption{Resistivity measurements of 200 nm Ce thin films upon cooling and warming.}
\label{res}
\end{figure*}
